\documentclass[aps,prfluids,groupedaddress,nofootinbib]{revtex4-1}

\usepackage{fullpage}
\usepackage{hyperref}
\hypersetup{
    colorlinks=true,
    linkcolor=blue,
    filecolor=magenta,      
    urlcolor=cyan,
    citecolor = {blue}
    }
\usepackage{amssymb,euscript,latexsym,amsmath}
\usepackage{graphicx}
\usepackage[dvips]{epsfig}
\usepackage{color}
\usepackage[dvipsnames]{xcolor}
\usepackage{epstopdf}
\usepackage{natbib}
\DeclareGraphicsExtensions{-eps-converted-to.pdf,.ps,.pdf}

\DeclareGraphicsExtensions{-eps-converted-to.pdf,.ps,.pdf}

\begin{document}


\title{Simulating cilia-driven mixing and transport in complex geometries}


\author{Hanliang Guo}
\author{Hai Zhu}
\author{Shravan Veerapaneni}
\email[]{shravan@umich.edu}
\affiliation{Department of Mathematics, University of Michigan, Ann Arbor, MI 48109}


\date{\today}

\begin{abstract}
Cilia and flagella are self-actuated microtubule-based structures that are present on many cell surfaces, ranging from the outer surface of single-cell organisms to the internal epithelial surfaces in larger animals. 
A fast and robust numerical method that simulates the coupled hydrodynamics of cilia and the constituent particles in the fluid such as rigid particles, drops or cells would be useful to not only understand several disease and developmental pathologies due to ciliary dysfunction but also to design microfluidic chips with ciliated cultures for some targeted functionality---e.g., maximizing fluid transport or particle mixing. In this paper, we develop a hybrid numerical method that employs a boundary integral method for handling the confining geometries and the constituent rigid particles and the method of regularized Stokeslets for handling the cilia. 
We provide several examples demonstrating the effects of confining geometries on cilia-generated fluid mixing as well as the cilia-particle hydrodynamics.
\end{abstract}


\maketitle


\section{Introduction}
Cilia are microscopic hair-like structures that protrude from cell surfaces. 
Motile cilia possess sophisticated internal structures, generally known as the ``9+2'' structures (see~\citet{Gibbons1981} for more details), that enable active periodic ciliary beatings. 
Being one of the most preserved structures in nature, cilia can be found in almost every phylum in the animal kingdom, from unicellular eukaryotes to invertebrate metazoans and vertebrates.
Cilia play significant roles in small animals such as unicellular organisms and invertebrate metazoans, including locomotion, generating water currents for feeding, and more~\cite{Gray1928, Sleigh1962, Machemer1985, vandenende1990}. In vertebrates, cilia are mostly found on the epithelial cell surfaces of internal organs, including the respiratory tract, brain, ear and oviduct~\cite{Afzelius1976, Hill1986, Verdugo1980, Fauci2006, Faubel2016, Olstad2019}, as the role in locomotion is replaced by muscles.

\textcolor{black}{The beating of individual cilium usually presents an asymmetric pattern that consists of a straight power stroke and a curly recovery stroke. This asymmetric pattern could break the famous ``scallop theorem'' (\citet{Purcell1977}) in viscous fluid at the individual level. Additionally, cilia are usually found in dense groups.} Each cell could feature hundreds of cilia in the respiratory tracts~\cite{Spassky2017}.
Interestingly, healthy cilia usually do not beat either in-phase or completely randomly. Instead, they beat in an orchestrated wavelike fashion: the so-called metachronal waves. Simply put, the metachronal waves are formed by all cilia performing similar beating patterns, but deforming in time with a small phase difference with respect to their neighbors.
The metachronal wave is an effective approach to transport fluid, \textcolor{black}{as it breaks the scallop theorem at the collective level.}

The study of ciliary/flagellar propulsion of micro-swimmers dates back to 1950s when G.I. Taylor~\cite{Taylor1951} modeled the flagellum of a sperm cell to an infinite sinusoidal traveling wave and studied analytically the relations between the swimming velocity and the wavenumber, the beating amplitude, and the traveling wave velocity.
We refer the reader to \citet{Lauga2009} and \citet{Gaffney2011} for a more detailed review on this topic. 
Ciliary transport in the airway systems has also received much attention due to its native relations with some human diseases~\cite{Fliegauf2007, Tilley2015}.
Numerous studies have been carried out using various numerical methods, including resistive force theory~\cite{Fulford1986}, slender body theory~\cite{Gueron1992, Gueron1997, Gueron1999}, immersed boundary method~\cite{Dillon2007, Lukens2010, Jayathilake2015}, immersed boundary-lattice Boltzmann method~\cite{Lee2011, Li2016, Chatelin2016}, finite element method~\cite{Mitran2007}, and the method of regularized stokeslet (MRS)~\cite{Smith2008a, Smith2008b, Smith2009d, Ding2014, Guo2014, Nawroth2017}.

\begin{figure}[!h]
        \centerline{\includegraphics[width=\linewidth]{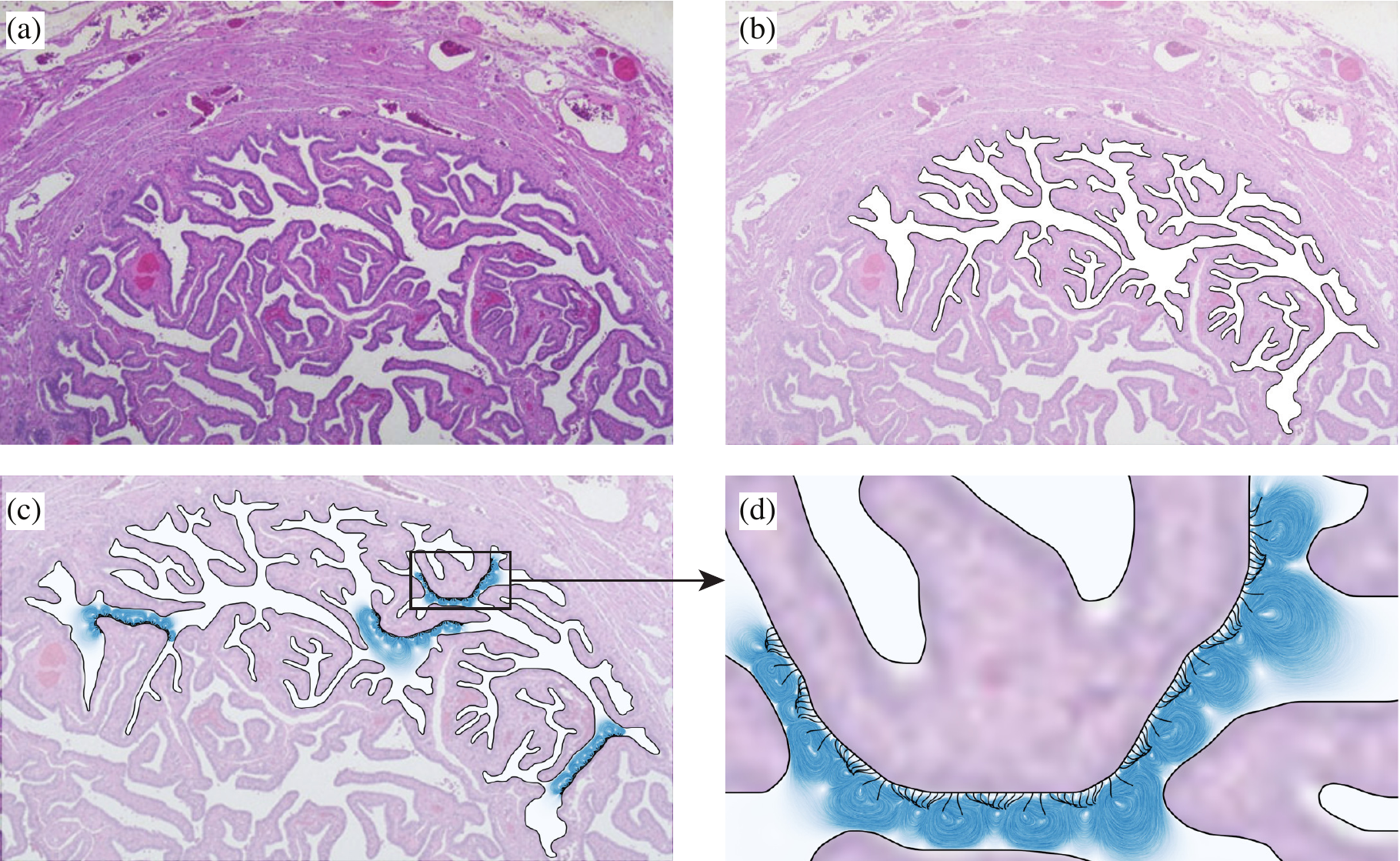}}
	\caption[]{An illustration of the current simulation capabilities of the hybrid numerical developed in this paper. For the purposes of this illustration, we  (a) took a generic microscopic image of the cross-section of a Fallopian tube (source: https://webpath.med.utah.edu/HISTHTML/NORMAL/NORM062.html), (b) extracted a subset of the fluid domain and its boundary highlighted here, (c) 
seeded four patches of the boundary with around four hundred cilia and solved the governing equations using the numerical method developed in this work and (d) visualized the solution via streamlines in one of the patches. 
\textcolor{black}{{\em Note:} This example is for illustrative purposes only and must not be viewed as representative of \textit{in vivo} flows. Physiologically, the major direction of fluid flow in the Fallopian tube is perpendicular to the cross-section as shown here; thereby, a three-dimensional simulation is needed to fully characterize the flows.} 
} \label{fig:fallopiantube}
\end{figure}

Ciliary mixing, on the other hand, has not been an active area of study until the last decade. 
Admittedly, mixing in viscous dominant fluid is inherently difficult due to the lack of turbulence~\cite{Ottino1989, Aref1990}.
Nevertheless,  \textit{in vivo} experiments of zebrafish embryo showed the transition of a unidirectional flow to vortical flows above and below the cilium tip~\cite{Supatto2008}, suggesting that the unidirectional transport is far from the sole purpose of the ciliary flow.
More recently, \citet{Nawroth2017} showed, using the squid-vibrio symbiotic systems, that long and short cilia that grow on the same ciliated organ serve different functions. Specifically, the long cilia beating with metachronal waves focus on the fluid transport and size-selective functions, while the short cilia with random phase differences enhance the fluid mixing with zero net flow on average.
Other works have also been able to show enhanced mixing using artificial cilia (e.g., see \cite{Fahrni2009, Shields2010, Chen2013, Saberi2019}).
Numerically, \citet{Lukens2010} studied the fluid mixing generated by a single cilium using the immersed boundary method and found distinct transport region and mixing region higher and lower than the cilium tip, respectively.
\citet{Ding2014} studied the fluid transport and mixing by a doubly-periodic array of cilia in a half-space bounded by a plane using MRS and its image systems. They systematically vary the phase differences between neighboring cilia and found consistent results in terms of the transport and mixing regions. Their results also showed that metachronal waves enhance not only fluid transport but also mixing. 
Recent works have also considered transport and mixing of multi-phase fluid in airway systems using the immersed boudary-lattice Boltzmann method~\cite{Chateau2018}. The results are qualitatively similar to that of \citet{Ding2014}, although tracers in different fluid layers are prevented from mixing  due to surface tension effects present at the interface.
More recently, \citet{Rostami2019} developed a large-scale simulation technique making using of the kernel-independent FMM and applied it on dense cilia carpets beating in phase; \citet{Stein2019} took a different route, wherein, instead of treating each cilium explicitly, they proposed an elegant coarse-grained model with anisotropic Brinkman equation and solved the cilia-driven transport problem using immersed boundary method. 

We note that all of the aforementioned computational works considered simple geometries and idealized boundary conditions such as periodic, free-space or half-space (bounded by a plane wall) flows. 
While such mathematical simplifications are important for problem tractability,  
the real environments that cilia beat in are far more complex.
For example, \citet{Faubel2016} showed that the complex flow of cerebrospinal fluid (CBF) in the delicate mice brain ventricles is regulated by the motile cilia; other human organs where cilia play important roles such as the tracheal and the Fallopian tubes present no less complex geometries~\cite{Bermbach2014, Lawrenson2013}.
Engineering applications such as manufacturing micro-fluidic devices that could transport and/or mix the fluid are also designed to have complex geometries (see, e.g.~\citet{Khaderi2011}).
To the best of our knowledge, no work to date has been focused on solving the cilia-driven flow in arbitrary complex geometries. Leveraging on recent advances in the boundary integral methods (BIMs) for Stokes flow, in this work, we develop methods for simulating  active cilia-driven flow of rigid particles in complex geometries. They are applicable in the regime where the hydrodynamics of the cilia-geometry and cilia-particle interactions are dominated by viscous effects and inertia could be neglected.


Specifically, we use the BIM for solving the Stokes equations inside the confining complex geometry and for evolving the rigid particles and the MRS for simulating the ciliary dynamics. Applying the boundary conditions at the fluid-structure interfaces leads to a set of coupled integro-differential equations at every time-step. We use the recently developed Nystr\"{o}m method in \citet{Wu2019} for discretizing the boundary integrals in these equations. This method is both $h-$adaptive and $p-$adaptive; that is, both the size of the boundary panels ($h$) and the degree of approximation per panel ($p$) are chosen automatically to achieve a prescribed error tolerance in the solution. Another advantage of this method is that the nearly singular integrals that arise due to the proximity of the walls, cilia and/or rigid particles are computed to high accuracy. A fourth-order explicit Runge-Kutta method is used for evolving both the cilia and the rigid particles. 

The current capabilities of this hybrid method are demonstrated by simulating the cilia-driven flow within the planar cross-section of a Fallopian tube as shown in figure~\ref{fig:fallopiantube}. Note that while the method can handle such arbitrary shapes, for simplicity and ease of analysis, the results section in this paper only considers flows in relatively classical geometries (e.g., Taylor-Couette channel, see figure~\ref{fig:schem}(a)). We emphasize here that such geometries are still difficult to handle using existing methods such as MRS with image systems.

The paper is organized as follows. We give the problem formulation and describe our numerical solvers in Section \ref{sc:methods}. Analysis of the mixing and trasport properties of actuated cilia in complex domains will be presented in Secton \ref{sc:results},  followed by conclusions and future work in Section \ref{sc:conclusions}.

\section{Model and methods}
\label{sc:methods}
In this section, we first describe the problem formulation for the specific case of cilia-driven flow of rigid particles suspended in a Taylor-Couette device. Then, we show how to recast it as a set of mixed boundary integral and discrete equations with unknowns residing on the cilia, particle and wall boundaries only (thus leading to dimensionality reduction). Lastly, we describe a numerical method for discretizing and solving these equations. 

\subsection{Model}
\begin{figure}[!h]
        \centerline{\includegraphics[width=.8\linewidth]{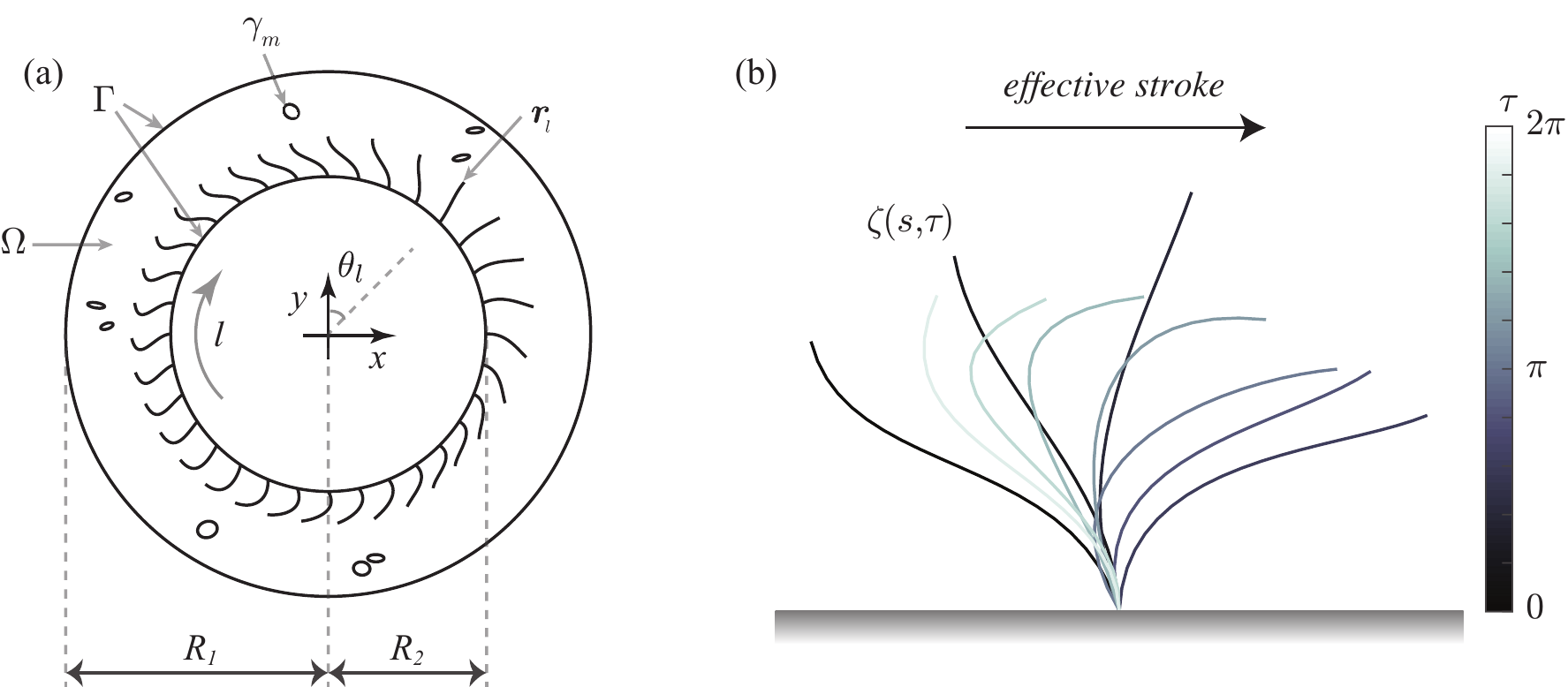}}
	\caption[]{Schematic figure. (a) $N$ cilia uniformly distributed at the inner surface of the stationary Taylor-Couette device. $N_p$ particles are freely suspended in the fluid domain $\Omega$ bounded by $\{(x,y)| R_2^2<x^2+y^2<R_1^2\}$.
	(b) The snapshots of the beating pattern extracted from \cite{Fulford1986}. Color-coded by its phase $\tau$.} 
	
	\label{fig:schem}
\end{figure}

Consider a thin gap of fluid confined between two stationary concentric circles of radius $R_1$ and $R_2$ with $R_1>R_2$. 
\textcolor{black}{The fluid domain is denoted by} $\Omega=\{(x,y)~|~ R_2^2<x^2+y^2<R_1^2 \}$.
$N_p$ rigid particles are immersed in the fluid; $N$ cilia of length $\ell$ are uniformly distributed on the surface of the inner circle~(see figure~\ref{fig:schem}(a)).
Following \cite{Fulford1986}, the kinematics $\boldsymbol{\zeta}$ of each cilium in its body frame can be approximated by a truncated Fourier series in time $\tau$ and a Taylor series in arc-length $s$.
The resulting beating pattern is shown in figure~\ref{fig:schem}(b).
We apply proper rotations and translations to $\boldsymbol{\zeta}$ to obtain the position of the $l$-th cilium $\boldsymbol{r}$ such that the cilia are uniformly distributed along the inner circle and are oriented perpendicular to the circle. 
Specifically, the position of the $l$-th cilium at arc-length $s$ and time $t$ is given by
\begin{equation}\label{eq:ciliapos}
\boldsymbol{r}_l(s,t) = 
\begin{bmatrix}
\cos(\theta_l) & \sin(\theta_l)\\
-\sin(\theta_l) & \cos(\theta_l)\\
\end{bmatrix}
\boldsymbol{\zeta}(s,\tau_l)+
R_2
\begin{pmatrix}
\sin(\theta_l)\\
\cos(\theta_l)\\
\end{pmatrix},
\quad
\theta_l = 2\pi(l-1)/N,
\quad
\tau_l = 2\pi t + (l-1)\Delta\phi.
\end{equation}
By construction, the 1st cilium is rooted at $(x,y)=(0,R_2)$ and the index of the cilium increases clockwise.
Here $\Delta\phi$ is a constant phase difference between neighboring cilia.
Specifically, all cilia beat in synchrony if $\Delta\phi = 0$; $0<\Delta\phi<\pi$ yields a wave that travels in the opposite direction as the effective stroke (antiplectic waves) and vice versa for $-\pi<\Delta\phi<0$ (simplectic waves).
\textcolor{black}{We note here that the stationary Taylor-Couette device is chosen to construct a ``periodic'' domain without the periodic kernels.}

In the low Reynolds number regime, the fluid dynamics is governed by the non-dimensional incompressible Stokes equation
\begin{equation}\label{eq:stokeseq}
-\nabla p(\boldsymbol{x}) + \nabla^2\boldsymbol{u}(\boldsymbol{x}) = \boldsymbol{0},\quad \nabla\cdot\boldsymbol{u} = 0, \quad \forall~\boldsymbol{x}\in\Omega.
\end{equation}
Here $p$ is the pressure, $\boldsymbol{u}$ is the velocity.
Taking advantage of the small aspect ratio of the cilia, we assume that the fluid velocity along the cilia is consistent with the prescribed beating pattern, namely
\begin{equation}\label{eq:bccilia}
\boldsymbol{u}(\boldsymbol{x}(s,t))=\frac{\mathrm{d} \boldsymbol{r}(s,t)}{\mathrm{d} t}, \quad \text{for } 0< s\le\ell.
\end{equation}
A no-slip boundary condition is applied on the wall boundaries, that is,
\begin{equation}\label{eq:bcboundary}
\boldsymbol{u}(\boldsymbol{x}) = \boldsymbol{0}, \quad \forall\,\, \boldsymbol{x}\in\Gamma. 
\end{equation}
On the other hand, given the translational velocity $\boldsymbol{U}_m$ and the angular velocity ${\omega}_m$ of the $m$-th rigid particle $(1\le m\le N_p)$, a no-slip boundary condition on $\gamma_m$ implies
\begin{equation}\label{eq:bcparticle}
\boldsymbol{u}(\boldsymbol{x}) = \boldsymbol{U}_m+ \omega_m (\boldsymbol{x}-\boldsymbol{x}_m^c)^{\perp}, \quad \forall \,\, \boldsymbol{x}\in\gamma_m,\\
\end{equation}
where $\boldsymbol{x}_m^c$ is the centroid of the particle and the {\em perp} operator $(\cdot)^\perp$ acts on vectors in $\mathbb{R}^2$ and is defined by \textcolor{black}{$\boldsymbol{x}^\perp = \begin{pmatrix} -x_2\\x_1\end{pmatrix}$}. 

Since the inertia is negligible in viscous dominant fluid, the particles need to also satisfy the no-net-force and no-net-torque conditions~\cite{Kim2013}. Particularly, in the absence of external forces and torques, we have
\begin{equation}\label{eq:nonetforce}
	\int_{\gamma_m} \boldsymbol{g} (\boldsymbol{x}) \, \mathrm{d}S_{\boldsymbol{x}} = \boldsymbol{0} \quad\text{and}\quad \int_{\gamma_m} \boldsymbol{g}(\boldsymbol{x})\cdot (\boldsymbol{x} - \boldsymbol{x}_m^c)^\perp \, \mathrm{d}S_{\boldsymbol{x}} = 0,
\end{equation}
where $\boldsymbol{g}$ denotes the hydrodynamic traction on the particle boundaries.

Throughout this paper, 
we normalize lengths by the typical length of cilia $\ell_c=20\mu\text{m}$, 
time by the typical beating period $T_c\approx 1/30 \text{s}$,
and force by $F_c={\mu \ell_c^2}/{T_c}=12\mathrm{pN}$, where $\mu = 10^{-3} \text{Pa}\cdot\text{s}$ is the water viscosity.

\subsection{Cilia-channel interactions}
We start by considering the simpler case where there are no rigid particles immersed in the fluid.
By virtue of linearity of the Stokes equation, the fluid velocity can be written as
\begin{equation}\label{eq:totalvel0}
\boldsymbol{u}(\boldsymbol{x}) = \boldsymbol{u}^c(\boldsymbol{x}) + \boldsymbol{u}^\Gamma(\boldsymbol{x}), \quad \forall \,\, \boldsymbol{x}\in\Omega,
\end{equation}
where $\boldsymbol{u}^c$ and $\boldsymbol{u}^\Gamma$ represent the disturbance flow due to the cilia and the boundary $\Gamma$ respectively.
We then take a hybrid approach: the governing equations for $\boldsymbol{u}^c$ are solved using MRS whereas those of $\boldsymbol{u}^\Gamma$ are solved using a BIM. 
Both approaches follow from our previous work in \cite{Guo2014} and \cite{Wu2019} respectively. 

We use an indirect integral equation formulation, also known as the combined field integral equation formulation \cite{Pozrikidis1992}, for the confining geometry $\Gamma$, which begins with an {\em ansatz} that the velocity is a sum of single and double layer potentials: 
\begin{equation}\label{eq:boundarydisturb}
	\boldsymbol{u}^\Gamma(\boldsymbol{x}) = 
	(\mathcal{S}_\Gamma+\mathcal{D}_\Gamma)[\boldsymbol{\mu}](\boldsymbol{x}), \quad \forall \,\, \boldsymbol{x}\in\Omega, 
\end{equation} 
where $\boldsymbol{\mu}$ is an unknown density function, $\mathcal{S}$ and $\mathcal{D}$ are the Stokes single- and double-layer operators respectively, defined as 
\begin{equation}\label{eq:potentials}
	\mathcal{S}_\Gamma[\boldsymbol{\mu}](\boldsymbol{x}) := \int_{\Gamma}\boldsymbol{G} (\boldsymbol{x},\boldsymbol{y})\boldsymbol{\mu}(\boldsymbol{y})\mathrm{d}S_{\boldsymbol{y}}\quad\text{and}\quad
		\mathcal{D}_\Gamma[\boldsymbol{\mu}](\boldsymbol{x}) := \int_{\Gamma}\boldsymbol{D}(\boldsymbol{x},\boldsymbol{y})\boldsymbol{\mu}(\boldsymbol{y})\,\mathrm{d}S_{\boldsymbol{y}}.
\end{equation} 
The convolution kernels $\boldsymbol{G}$ and $\boldsymbol{D}$ are the fundamental solutions to Stokes equations \eqref{eq:stokeseq}, written in component form as, 
\begin{equation}\label{eq:kernel}
\begin{split}
	G_{ij}(\boldsymbol{x},\boldsymbol{y}) &= \frac{1}{4\pi}\left(\delta_{ij}\log\frac{1}{|\boldsymbol{x}-\boldsymbol{y}|} + \dfrac{(x_i-y_i)(x_j-y_j)}{|\boldsymbol{x}-\boldsymbol{y}|^2}\right), \\
	D_{ij}(\boldsymbol{x},\boldsymbol{y}) &= \frac{1}{\pi}\left(\dfrac{n_k(\boldsymbol{y})(x_k-y_k)}{|\boldsymbol{x}-\boldsymbol{y}|^2} \dfrac{(x_i-y_i)(x_j-y_j)}{|\boldsymbol{x}-\boldsymbol{y}|^2} \right),
\end{split}
\end{equation} 
where $\boldsymbol{n}(\boldsymbol{y})$ is the outer normal vector of the surface $\Gamma$ at $\boldsymbol{y}$. By definition, \eqref{eq:boundarydisturb} satisfies the Stokes equations and what remains is to enforce the no-slip boundary condition \eqref{eq:bcboundary}. 
\textcolor{black}{To simplify the formulation, we reverse the order of the discretization points on the inner circle so that the fluid domain can be considered ``interior'' to both outer and inner circles.}
Taking the limit as the target $\boldsymbol{x}$ approaches $\Gamma$ from the interior 
and using standard jump conditions for the layer potentials \textcolor{black}{(e.g., see \citet{ladyzhenskaya1969mathematical}, Chapter 3)}, we arrive at the following equation: 
\begin{equation} \label{eq:IE}
 \left(-\frac{1}{2} I + \mathcal{S}_\Gamma + \mathcal{D}_\Gamma\right)[\boldsymbol{\mu}](\boldsymbol{x})  = - \boldsymbol{u}^c (\boldsymbol{x}), \quad \forall \,\, \boldsymbol{x}\in \Gamma.
\end{equation}
The above is a second-kind integral equation (SKIE) for the unknown density function $\boldsymbol{\mu}$. The main advantage of SKIEs is that they result in a well-conditioned linear system when discretized.
\textcolor{black}{However, this system has a null-space of dimension one in the direction normal to the boundary, which needs to be eliminated via standard techniques (e.g., see \cite{sifuentes2015randomized}). Particularly, we eliminate the null-space by adding the components of the normal vector of each quadrature point in the first column of the resulted discretization matrix.}
A more popular approach for obtaining SKIE in this context is the completed double-layer formulation of Power and Miranda \cite{power1987second}. We chose the above for simplicity (we will employ the same formulation for particles as well). Note, however, that the right hand side vector $\boldsymbol{u}^c$ in \eqref{eq:IE} is also unknown; we discuss its formulation next.    

We discretize each cilium into $N_s$ uniformly placed beads along the arc-length.
Specifically, the position of the $m$-th bead for one cilium is at arclength $s=m\ell/N_s$ with $1\le m\le N_s$.
We treat each bead along the cilia as a 2D free-space regularized force. The flow field can then be reconstructed using the method of regularized stokeslet~\cite{Cortez2001}.
Assuming the position and strength of the $n$-th regularized forces are $\boldsymbol{r}_n$ and $\boldsymbol{f}_n$, the flow field generated by such a force distribution is given by 
\begin{equation}\label{eq:ciliadisturb}
\boldsymbol{u}^c (\boldsymbol{x}) = \sum_{n=1}^{N N_s} \widetilde{\boldsymbol{G}}(\boldsymbol{x},\boldsymbol{r}_n) \boldsymbol{f}_n,
\end{equation}
where $\widetilde{\boldsymbol{G}}$ is the regularized version of the Green's function $\boldsymbol{G}$ given in \eqref{eq:kernel}, defined in the Appendix \ref{sc:regularized}.
From here on, a regularized operator will be denoted using the symbol ~$\widetilde{}$~ as above. Since the representation \eqref{eq:ciliadisturb} can be viewed as a discrete, regularized analogue of the layer potentials \eqref{eq:potentials}, we will use a similar notation; that is, we let
\begin{equation}\label{eq:ciliaS}
\widetilde{\mathcal{S}}_c[\boldsymbol{f}](\boldsymbol{x})  := \sum_{n=1}^{N  N_s} \widetilde{\boldsymbol{G}}(\boldsymbol{x},\boldsymbol{r}_n) \boldsymbol{f}_n.
\end{equation}
Since the cilia beating pattern is assumed to be given {\em a priori}, the force density $\boldsymbol{f}$ is unknown. This will be determined by enforcing the no-slip boundary condition at the cilia-fluid interface \eqref{eq:bccilia}. Together with \eqref{eq:IE}, the system of equations for the two unknown densities can then be summarized in the matrix form as
\begin{equation}\label{eq:totalvel_mat}
\begin{bmatrix}
-\frac{1}{2}I + \mathcal{S}_{\Gamma, \Gamma}+\mathcal{D}_{\Gamma, \Gamma} & \widetilde{\mathcal{S}}_{c, \Gamma}\\
\mathcal{S}_{\Gamma, c}+\mathcal{D}_{\Gamma, c} & \widetilde{ \mathcal{S}}_{c, c}\\
\end{bmatrix}
\begin{bmatrix}
\boldsymbol{\mu}\\
\boldsymbol{f}
\end{bmatrix}
=\begin{bmatrix}
\boldsymbol{0}\\
\frac{\mathrm{d}\boldsymbol{r}}{\mathrm{d}t}
\end{bmatrix}.
\end{equation}
Here, with a slight abuse of notation, we denoted the single-layer potential defined on $\Gamma$ (sources) evaluated at the discrete points on the cilia (targets) by  $\mathcal{S}_{\Gamma, c}$. Other operators are defined analogously.  Once we solve this matrix equation, we can evaluate the velocity field at any point $\boldsymbol{x}$ in the fluid domain by using \eqref{eq:totalvel0}.

Notice that the operator $\left(-\frac{1}{2}I + \mathcal{S}_{\Gamma, \Gamma} + \mathcal{D}_{\Gamma, \Gamma}\right)$ in the matrix equation \eqref{eq:totalvel_mat} remains fixed as the cilia beat, since $\Gamma$ is stationary. Therefore, it is computationally efficient to simply compute its inverse (once discretized) as a precomputation step before time-stepping for the evolution of cilia. In large-scale systems (such as in Figure \ref{fig:fallopiantube}), one can accelerate this precomputation, and application of inverse at every time-step, using a low-rank factorization based fast direct solver as done recently for similar problems in \cite{marple2016fast}.

\subsection{Cilia-channel-particle interactions}
We now extend our formulation to include rigid particles suspended in ciliary-driven flow in confining geometries. For notational simplicity, we consider the case where only a single rigid particle with boundary $\gamma$ is present. In this case, the velocity in the fluid domain can be decomposed into three components due to disturbance flows created by the cilia, the stationary wall and the particle respectively as
\begin{equation}\label{eq:totalvel2}
\boldsymbol{u} (\boldsymbol{x}) =\boldsymbol{u}^{c}(\boldsymbol{x})+\boldsymbol{u}^\Gamma (\boldsymbol{x})+\boldsymbol{u}^\gamma (\boldsymbol{x}), \quad \forall\, \boldsymbol{x}\in \Omega.
\end{equation}

Similar to earlier treatment, $\boldsymbol{u}^{c}$ and $\boldsymbol{u}^{\Gamma}$ are given by \eqref{eq:ciliadisturb} and \eqref{eq:boundarydisturb} respectively. For $\boldsymbol{u}^\gamma$, we use the same ansatz as in \eqref{eq:boundarydisturb}, that is, we write:
\begin{equation}\label{eq:particldisturb}
	\boldsymbol{u}^{\gamma}(\boldsymbol{x}) =
	(\mathcal{S}_\gamma+\mathcal{D}_\gamma)[\boldsymbol{\mu}](\boldsymbol{x}), \quad \forall\, \boldsymbol{x}\in\Omega. 
\end{equation} 
The vector density function $\boldsymbol{\mu}$ defined on $\gamma$ again is an unknown that needs to be determined by applying the boundary conditions on the particle boundaries.  By taking the limit of \eqref{eq:particldisturb} as $\boldsymbol{x}$ approaches $\gamma$ from the exterior and applying the rigid body velocity condition \eqref{eq:bcparticle} yields the following BIE,
\begin{equation}\label{eq:paricleBIE}
\left(\frac{1}{2} I + \mathcal{S}_{\gamma} + \mathcal{D}_{\gamma}\right)[\boldsymbol{\mu}] (\boldsymbol{x}) =  - \boldsymbol{u}^c(\boldsymbol{x}) - \boldsymbol{u}^\Gamma (\boldsymbol{x}) + \boldsymbol{U} + \omega (\boldsymbol{x} - \boldsymbol{x}^c)^\perp, \quad \forall \,\, \boldsymbol{x} \in \gamma.
\end{equation}
Again, the above is a SKIE for the unknown $\boldsymbol{\mu}$ defined on $\gamma$. The rigid body translational and rotational velocities $(\boldsymbol{U}, \omega)$ are also unknown {\em a priori} and need to be solved for by applying the force- and torque-free conditions \eqref{eq:nonetforce} on $\gamma$. To do so, we need to evaluate the hydrodynamic traction on $\gamma$ based on the velocity representation \eqref{eq:totalvel2} which can now be written in its full form as
\begin{equation}\label{eq:veldecomp}
	\boldsymbol{u}(\boldsymbol{x}) = \widetilde{\mathcal{S}}_c[\boldsymbol{f}](\boldsymbol{x})  + 
	(\mathcal{S}_\Gamma+\mathcal{D}_\Gamma)[\boldsymbol{\mu}](\boldsymbol{x}) + (\mathcal{S}_\gamma+\mathcal{D}_\gamma)[\boldsymbol{\mu}](\boldsymbol{x}). 
\end{equation} 
The traction force at $\gamma$ can be computed using the formula $\, \boldsymbol{g}(\boldsymbol{x}) = -p(\boldsymbol{x})\boldsymbol{n}(\boldsymbol{x}) + (\nabla \boldsymbol{u}(\boldsymbol{x}) + \nabla \boldsymbol{u}^T(\boldsymbol{x}))\cdot\boldsymbol{n}(\boldsymbol{x}), \,$ where $\boldsymbol{n}$ is the outward normal to $\gamma$. 
We can avoid computing the derivatives numerically by plugging \eqref{eq:veldecomp} into this formula and evaluating the derivatives of the integral kernels analytically. For notational convenience, we define the velocity vector $\boldsymbol{U}^*$ and the operators $\mathcal{L}$ and $\mathcal{C}$ as
\begin{equation}\label{eq:particleops}
\boldsymbol{U}^* = \begin{bmatrix} \boldsymbol{U} \\ \omega \end{bmatrix}, \qquad
\mathcal{L}_\gamma \, \boldsymbol{g} := 
\begin{bmatrix} 
\int_\gamma \boldsymbol{g} (\boldsymbol{x})\mathrm{d}S_{\boldsymbol{x}} \\
\int_\gamma \boldsymbol{g}(\boldsymbol{x})\cdot(\boldsymbol{x} - \boldsymbol{x}^c)^\perp \mathrm{d}S_{\boldsymbol{x}} 
\end{bmatrix} \quad\text{and}\quad \mathcal{C} \, \boldsymbol{U}^* := \boldsymbol{U} + \omega  (\boldsymbol{x} - \boldsymbol{x}^c)^\perp. 
\end{equation}
Since the force- and torque-free conditions \eqref{eq:nonetforce} does not require the pointwise values for the traction, we can utilize the identity that the action of $\mathcal{L}_\gamma$ on the Stokes double-layer potentials in \eqref{eq:veldecomp} produces the zero vector (e.g., see \cite{Pozrikidis1992, barnett2018unified}). What remains is to evaluate the traction force due to the single-layer potentials in \eqref{eq:veldecomp}. The traction associated to the single layer potential $\mathcal{S}_\gamma[\boldsymbol{\mu}]$, for example, is given by 
\begin{equation} \label{eq:traction}
\mathcal{K}_\gamma[\boldsymbol{\mu}](\boldsymbol{x})_i := \int_{\gamma}T_{ijk}(\boldsymbol{x},\boldsymbol{y}) n_k(\boldsymbol{x}) \mu_j(\boldsymbol{y})\mathrm{d}S_{\boldsymbol{y}},
\end{equation}
where the traction kernel, also known as the stresslet, is given by 
\begin{equation}
T_{ijk}(\boldsymbol{x},\boldsymbol{y}) = -\frac{1}{\pi}\frac{(x_i-y_i)(x_j-y_j)(x_k-y_k)}{|\boldsymbol{x}-\boldsymbol{y}|^4}.
\end{equation}
The regularized traction kernel is given in Appendix \ref{sc:regularized}. Based on these definitions, we can now write the force- and torque-free conditions on the rigid particles as 
\begin{equation}\label{eq:traction0}
\mathcal{L}_\gamma\left( - \frac{1}{2} \boldsymbol{\mu}(\boldsymbol{x}) + \mathcal{K}_\gamma[\boldsymbol{\mu}](\boldsymbol{x}) + \mathcal{K}_\Gamma[\boldsymbol{\mu}](\boldsymbol{x}) +\widetilde{\mathcal{K}}_c[\boldsymbol{f}](\boldsymbol{x})\right) = \boldsymbol{0}, \quad \forall \,\, \boldsymbol{x} \in \gamma. 
\end{equation}
Therefore, together with \eqref{eq:paricleBIE}, this equation is sufficient to determine the unknowns $\boldsymbol{\mu}$ and $\boldsymbol{U}^*$ residing on $\gamma$. The coupled system of equations for all the unknowns can now be summarized in the matrix form as
\renewcommand{\arraystretch}{1.5}
\begin{equation}\label{eq:fullsys}
\begin{bmatrix}
-\frac{1}{2}I + \mathcal{S}_{\Gamma,\Gamma}+\mathcal{D}_{\Gamma,\Gamma} & \mathcal{S}_{\gamma, \Gamma}+\mathcal{D}_{\gamma,\Gamma} &  \widetilde{\mathcal{S}}_{c,\Gamma} & 0\\
\mathcal{S}_{\Gamma,\gamma}+\mathcal{D}_{\Gamma,\gamma} & \frac{1}{2}I + \mathcal{S}_{\gamma,\gamma}+\mathcal{D}_{\gamma,\gamma} &  \widetilde{\mathcal{S}}_{c,\gamma} & -\mathcal{C} \\
\mathcal{S}_{\Gamma, c} + \mathcal{D}_{\Gamma, c} & \mathcal{S}_{\gamma, c} + \mathcal{D}_{\gamma,c} &  \widetilde{\mathcal{S}}_{c,c} & 0\\
\mathcal{L}_\gamma \, \mathcal{K}_{\Gamma, \gamma} & \mathcal{L}_\gamma \left(-\frac{1}{2} I + \mathcal{K}_{\gamma, \gamma} \right) &  \mathcal{L}_\gamma\, \widetilde{\mathcal{K}}_{c, \gamma} & 0
\end{bmatrix}
\begin{bmatrix}
\boldsymbol{\mu}(\Gamma)\\
\boldsymbol{\mu}(\gamma) \\
\boldsymbol{f}\\
\boldsymbol{U}^*
\end{bmatrix}
=
\begin{bmatrix}
 \boldsymbol{0}\\
 \boldsymbol{0}\\
\dfrac{d\boldsymbol{r}}{dt}\\
\boldsymbol{0}
\end{bmatrix}.
\end{equation}
While the above matrix form is helpful in understanding the overall formulation, in practice, we may invert smaller systems depending on the nature of the problem. For example, as discussed earlier, if the number of unknowns on $\Gamma$ is large compared to others, it would be beneficial to form its inverse as a precomputation step. All the layer potentials in this system matrix lead to $N-$body sums when discretized, thereby, require fast algorithms to accelerate their computation for large problem sizes. Many such algorithms are now well-established; we use the open-source fast multipole method (FMM) implementation of  \cite{HFMM2D}. Note that solving this system gives all the unknowns at a particular time snapshot only; we then have to update the position of cilia using \eqref{eq:ciliapos} and the position of the rigid particle using $\boldsymbol{U}^*$.

Finally, the formulation generalizes to multiple rigid particles in a trivial manner. We let $\gamma$ denote the union of all the particle boundaries i.e., $\gamma = \bigcup_{m=1}^{N_p} \gamma_m$, where, as before, $\gamma_m$ is the boundary of the $m$-th particle. Then, the definition of the boundary integral operators introduced so far hold as is; for example, 
\begin{equation}
\mathcal{S}_\gamma[\boldsymbol{\mu}](\boldsymbol{x}) = \int_{\gamma}\boldsymbol{G} (\boldsymbol{x},\boldsymbol{y})\boldsymbol{\mu}(\boldsymbol{y})\mathrm{d}S_{\boldsymbol{y}}
 := \sum_{m = 1}^{N_p} \int_{\gamma_m} \boldsymbol{G} (\boldsymbol{x},\boldsymbol{y})\boldsymbol{\mu}(\boldsymbol{y})\mathrm{d}S_{\boldsymbol{y}}.
\end{equation}
The operators and variables in \eqref{eq:particleops}, on the other hand, must be defined separately for each particle and the system \eqref{eq:fullsys} needs to be modified accordingly by concatenating the unknowns on all the particle boundaries. 

\subsection{Nystr\"om discretization and close-evaluation of layer potentials}
Given a single closed curve $\Gamma$ parameterized by $\boldsymbol{Z}(\alpha): [0,2\pi) \to \mathbb{R}^2$, such that $\Gamma = \boldsymbol{Z}([0,2\pi))$, we split the curve uniformly into $n_\Lambda$ disjoint panels $\Lambda_i$, $i = 1,\ldots,n_\Lambda$. In each panel, we use $p$ quadrature nodes so that there are $N_{\Gamma}=pn_\Lambda$ discrete points on the curve. 
The standard Gauss-Legendre quadrature, with nodes $\{t_i\}_{i=1}^{N_{\Gamma}}$ and associated weights $\{w_i\}_{i=1}^{N_{\Gamma}}$, 
offers high-order accuracy for integrating any smooth function $g$ on $\Gamma$, 
	\begin{equation} \label{eq:plain}
		\int_\Gamma g(\boldsymbol{y})\mathrm{d}S_{\boldsymbol{y}}  = \int_0^{2\pi} g(\boldsymbol{Z}(\alpha)) \,|\boldsymbol{Z}'(\alpha)|\, d\alpha \; \approx \; 
\sum_{i=1}^{N_\Gamma} g(\boldsymbol{Z}(\alpha_i))\, |\boldsymbol{Z}'(\alpha_i)|\, w_i.
	\end{equation}

Now consider the task of computing the velocity $\boldsymbol{u}(\boldsymbol{x})$ at a target $\boldsymbol{x}\in\Omega$ by evaluating \eqref{eq:boundarydisturb}. 
If $\boldsymbol{x}$ is far away from a source panel $\Lambda$, the contribution of $\Lambda$ to $\boldsymbol{u}$ is obtained by simply using the quadrature rule \eqref{eq:plain} since the integrand is smooth in this case.
However if $\boldsymbol{x}$ is close to $\Lambda$, one could expect the integral kernels in \eqref{eq:potentials} to be much more rapidly changing functions of $\boldsymbol{y}\in\Gamma$ than $\boldsymbol{\mu}$. In fact, the error in a fixed smooth quadrature rule grows exponentially to $\mathcal{O}(1)$ as $\boldsymbol{x}$ approaches $\Gamma$. These inaccuracies may lead to numerical instabilities. Therefore we adapt a local panelwise close evaluation scheme proposed in \cite[Sec.~3]{Wu2019} to accurately handle nearly singular hydrodynamic interactions. This is done by first rewriting velocity field $\boldsymbol{u}$ represented by Stokes single or double layer potential in terms of complex contour integrals with different types of singularity and then using a high-order polynomial interpolation in complex plane to approximate the density function $\boldsymbol{\mu}$. We may integrate analytically the resulting contour integral of each monomial using a two-term recurrence. This specialized panel quadrature scheme provides uniform accuracy for targets arbitrarily close to, or on, the curve.

\section{Results and discussions}
\label{sc:results}
The numerical parameters used in this section are listed in Table~\ref{tab:parameters}.
\begin{table}
\begin{center}
\begin{tabular}{ccc}
\hline
Parameter & Symbol &  non-dimensional value \\
\hline
Number of cilia & $N$ & 32 \\
Number of regularized stokeslets per cilium & $N_s$ & 20 \\
Regularization parameter & $\epsilon$ & $1/80$ \\
Radius of the outer wall & $R_1$ & 5\\
Radius of the inner wall & $R_2$ & 3\\
Quadrature points on the channel walls & & $\approx3000$ \\
Quadrature points on each particle & & 128 \\
 Panel order & & 16 \\
Number of waves & $N_w$ & $-10,-9,-8,\cdots, 10$ \\
Phase difference & $\Delta\phi$ & $\frac{\pi}{16}N_w$\\
Time step& $\Delta t$ & 1/200\\
\hline
\end{tabular}
\caption{List of numerical parameters.}\label{tab:parameters}
\end{center}
\end{table}
With these choices of the parameters, we are able to achieve a close to machine-precision accuracy with the spatial scheme and a forth order convergence with the temporal scheme.
The numerical validations are shown in the Appendix.

\subsection{Mixing of tracers}
\begin{figure}[!h]
        \centerline{\includegraphics[width=.9\linewidth]{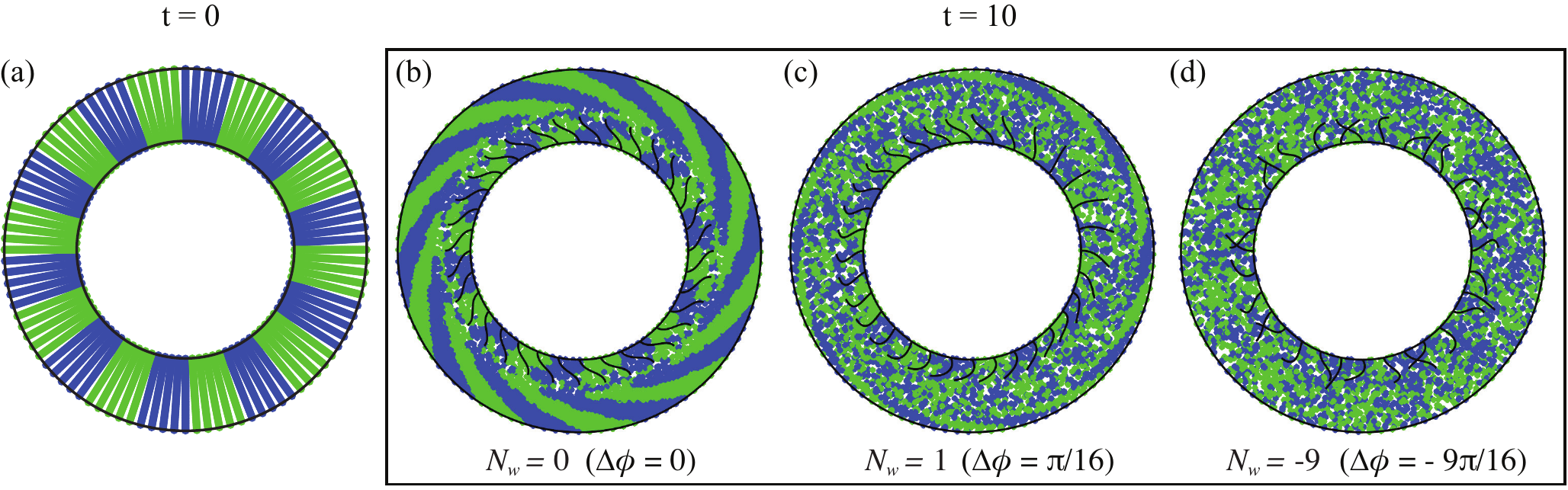}}
	\caption[]{Uniformly seeded tracers mixed by cilia with different phase differences after 10 beating cycles. (a) Initial seeding; (b) $N_w=0$; (c) $N_w=1$; (d) $N_w=-9$.} 
	
	\label{fig:mixingresults}
\end{figure}

We first apply the numerical method to study the mixing of passive tracers.
Specifically, we uniformly seed 5000 tracers inside the channel and color them blue or green as shown in figure~\ref{fig:mixingresults}(a). We track the tracers for 10 beating cycles and visualize the positions of the passive tracers in figure~\ref{fig:mixingresults}(b-d) for three different phase differences between neighboring cilia. 
We use the number of waves formed by the cilia $N_w$ as a proxy of the phase difference $\Delta\phi$ to make sure that there are always complete waves in the channel. One can convert between the two variables using the relation $\Delta\phi = \frac{\pi}{16}N_w$.
The waves travel in the counter-clockwise direction if $N_w>0$, which corresponds to the antiplectic metachronal waves, and vice versa for $N_w<0$, which corresponds to the symplectic metachronal waves.
Clearly, in the case of cilia beating in synchrony ($N_w=0$), the tracers are barely mixed - a shear region could be  identified between the tips of the cilia and the outer channel wall, consistent with previous numerical results~\cite{Lukens2010, Ding2014}. 
Note that although all the cilia are beating in synchrony, the asymmetry between the effective and recovery strokes drives the flow over one cycle.
The mixing performance becomes much stronger as the phase difference becomes non-zero. Two representative cases are shown in figure~\ref{fig:mixingresults}(c)\&(d) with $N_w=1$ and $N_w=-9$. 
In the case of $N_w=1$, the mixing region becomes much larger compare to the synchronized case while a small shear region is still observable close to the outer channel wall. On the other hand, the mixing region completely dominates the shear region when $N_w=-9$. Note that the diminished shear region and the mixing region above the ciliary tips have not been reported before in other geometries. 
The authors speculate that this is due to the narrowness of the channel and the no-slip boundary condition on the channel walls. 
\textcolor{black}{Additionally, the fact that the fluid domain is closed and there is a lack of ``fresh'' supply of fluid could also be a reason for the diminished shear region.}

To quantify the mixing performance, following~\cite{Stone2005}, we use the shortest distance between particles of different colors as a measure. 
Let $N_t$ be the total number of tracers of each color, $i,j$ be the indices of the blue and green tracers respectively. The mixing number could be defined as 
\begin{equation}
\mathrm{m} = \left( \prod_{i=1}^{N_t} \min_j(|\boldsymbol{x}_i - \boldsymbol{x}_j|)^2\right)^{1/N_t}.
\end{equation}
Note that the mixing number $\mathrm{m}$ is positive by definition; a well mixed state has a mixing number close to 0.

\begin{figure}[!h]
        \centerline{\includegraphics[width=.85\linewidth]{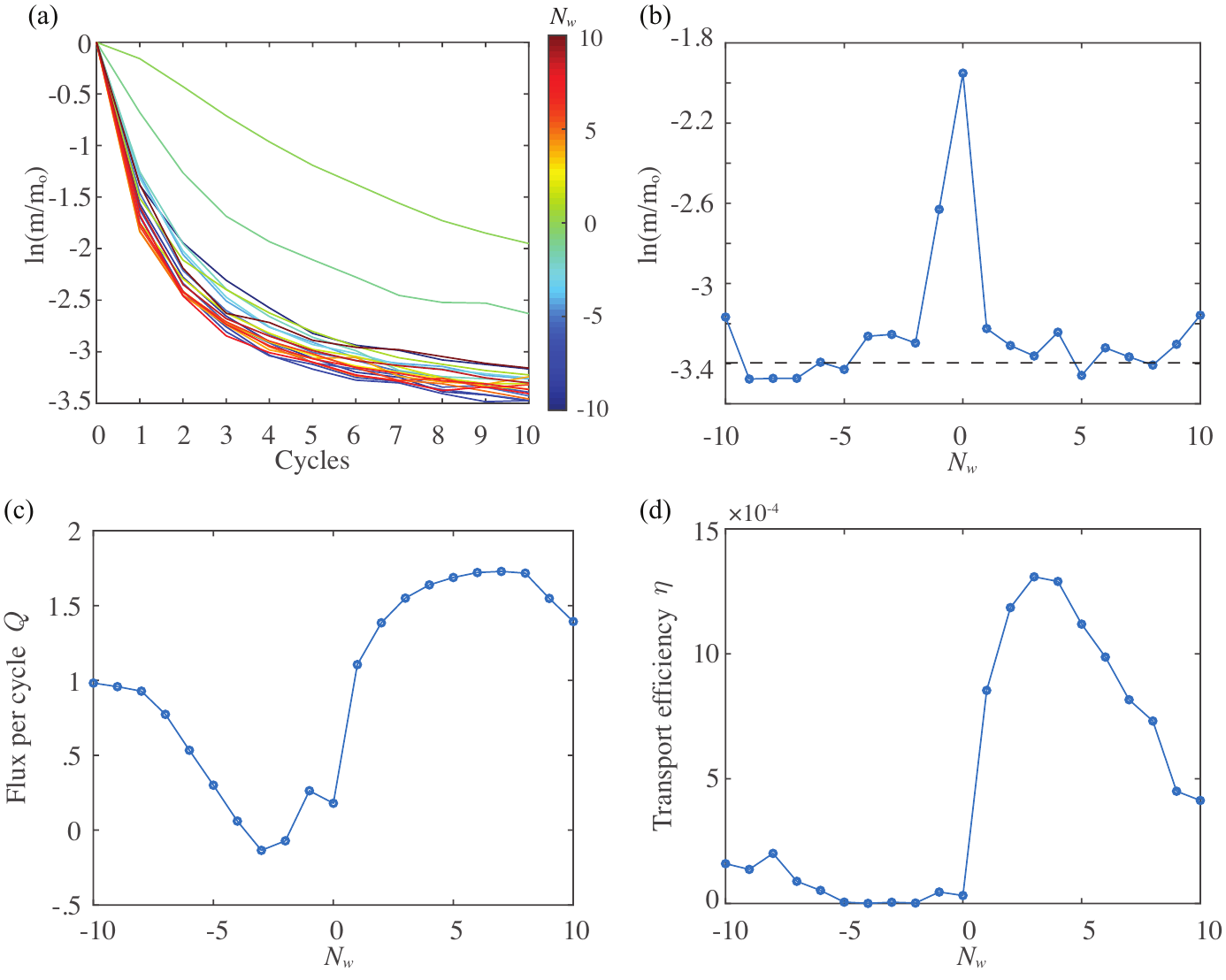}}
	\caption[]{Mixing and transport performance. (a) Mixing efficiency as a function of cycles for different number of waves (phase difference). (b) Mixing efficiency after 10 cycles as a function of the number of waves $N_w$. (c) Total flux per cycle as a function of $N_w$. \textcolor{black}{(d) Transport efficiency as a function of $N_w$.}} 
\label{fig:mixingandtransport}
\end{figure}
In figure~\ref{fig:mixingandtransport}(a) we show the mixing number normalized by the initial mixing number $\mathrm{m}_o=\mathrm{m}|_{t=0}$ as a function of beating cycles in semi-logarithm axes.
A wide range of phase differences are depicted by the different line colors.
The mixing numbers decrease fast in the first couple of cycles and start to plateau afterwards.
For all phase differences, the mixing number decreases as a function of cycles, indicating that the ciliary beating keeps mixing the fluid.
The case that all cilia beat in synchrony ($N_w=0$) has the worst mixing performance, consistent with the ``scallop theorem''~\cite{Purcell1977}.
The mixing numbers after 10 beating cycles are shown in figure~\ref{fig:mixingandtransport}(b) as a function of number of waves $N_w$. 
The case that yields the best mixing performance is $N_w=-9$, corresponds to a phase difference $\Delta\phi=-9\pi/16$.
It is also evident that the mixing performance is relatively robust to the phase difference.
In fact, almost all mixing numbers reach as low as $e^{-3.4}\approx0.03$ after normalization except for the two cases where $N_w = 0$ and $-1$ ($\Delta\phi=0$ and $-\pi/16$).
Note that this is in contrast to what has been observed previously in~\cite{Ding2014} for an idealized geometry where they found the mixing performances are sensitive to the phase differences and two clear local extrema were identified.

We continue by examining the transport performance of the ciliary flow in the channel.
We quantify the transport by evaluating the total flux \textcolor{black}{$Q$} going through a vertical cross-section $\{(x,y)|x=0, R_2<y<R_1\}$ over one beating cycle.
To be consistent with the ciliary effective stroke direction, we take the positive $x$-direction as the positive direction for the flux at this cross-section.
By virtue of incompressibility, the flux going through different cross-sections are equal to each other, which we verified in our simulations (results not shown here).
The total flux per cycle is shown in figure~\ref{fig:mixingandtransport}(c) as a function of phase differences.
Similar to the mixing performance, having all cilia beating in synchrony generates almost no transport due to the scallop theorem.
Additionally, antiplectic waves ($N_w>0$) in general perform better than simplectic waves ($N_w<0$).
The case that generates the largest flux is $N_w=7$ ($\Delta\phi=7\pi/16\approx0.44\pi$), which is similar to what authors in \cite{Ding2014} report, albeit in different geometries.
\textcolor{black}{To determine the transport efficiency, we follow the previous works of \citet{Osterman2011, Eloy2012, elgeti2013emergence} and \citet{Guo2014}, and define the dimensionless transport efficiency $\eta$ as
\begin{equation}\label{eq:efficiency}
\eta = \mu \ell^{-1}\frac{ Q^2}{W},
\end{equation}
where $W = \frac{1}{N_c}\sum_i^{Nc} \int_0^T\int_0^\ell \max(0,\boldsymbol{q}\cdot\boldsymbol\alpha) \mathrm{d}s\mathrm{d}t$ is the power loss over one beating cycle averaged over per cilium, $\boldsymbol\alpha=\|\dot{\boldsymbol{t}(s)}\|\frac{\boldsymbol{t}\times\dot{\boldsymbol{t}}}{\|\boldsymbol{t}\times\dot{\boldsymbol{t}}\|}$ is the angular velocity vector, $\boldsymbol{q} = \boldsymbol{t}''\times\boldsymbol{t}+\boldsymbol{t}\times\int_s^\ell \boldsymbol{f}(\tilde{s},t)d\tilde{s}$ is the internal moments generated along each cilium, and only positive works are accounted for. The results show that the phase difference that optimizes transport efficiency, $\Delta\phi = 3\pi/16\approx0.19\pi$, is smaller compare to that optimizes the total flux, due to the high power loss at larger phase differences (power loss results not shown here). Overall, the transport efficiency in viscous fluid remains small, consistent with previous works.}

\begin{figure}[!h]
        \centerline{\includegraphics[width=.9\linewidth]{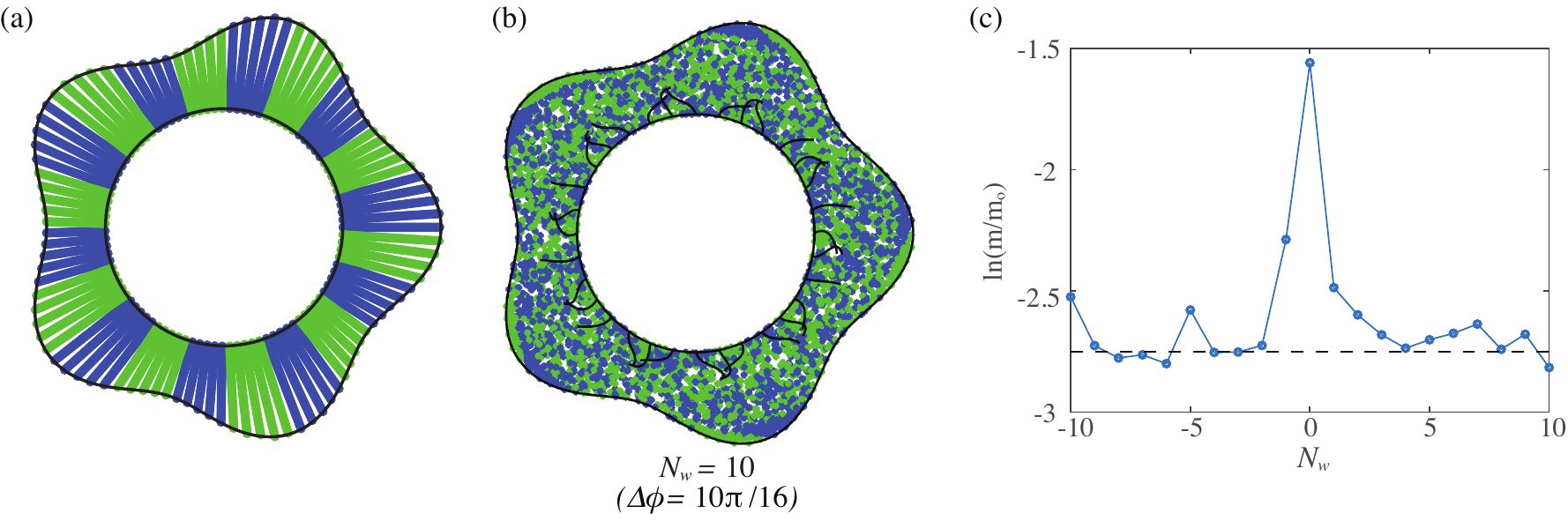}}
	\caption[]{Uniformly seeded tracers mixed by cilia in a wavy channel. (a) Initial seeding. (b) Tracers after 10 cycles for $N_w=10$ ($\Delta\phi = 10\pi/16$). (c) Mixing efficiency after 10 cycles as a function of $N_w$.} 	
	\label{fig:mixingresults2}
\end{figure}

To further illustrate the effects of geometries on the mixing performance, we study the mixing results of ciliary beating inside a ``wavy channel''.
Specifically, we perturb the outer channel wall such that the boundary can be written as $z=x+\mathrm{i}y=\frac{R_1(1+0.1\cos(5\theta))}{\sqrt{1+0.1^2/2}}\exp(\mathrm{i}\theta)$, $\theta\in[0,2\pi)$ in complex form. 
The coefficient in the denominator is to scale the channel such that the fluid domain has the same \textcolor{black}{area} compared to the regular circular channels.
\textcolor{black}{The coefficient of the cosine term perturbs the radius of the outer boundary by about $\pm10\%$. In other words, the narrowest and the widest channel widths are about $1.5$ and $2.5$ unit length.}
The initial seeding and the tracer positions after 10 beating cycles are shown in figure~\ref{fig:mixingresults2}(a)\&(b) with $N_w=10$, which yields the best mixing results as shown in~figure~\ref{fig:mixingresults2}(c).
When compared to figure~\ref{fig:mixingandtransport}, it is clear that not only the number of waves that yields the best mixing performance changes from $-9$ to $10$, but also the overall mixing performance is negatively affected by the presence of  the wavy channel -- indicated by the mixing number $\ln(\mathrm{m}/\mathrm{m_o})$ increased from $-3.4$ to about $-2.8$ (in other words, $\mathrm{m}/\mathrm{m_o}$ increased from $e^{-3.4}\approx 0.03$ to $e^{-2.8}\approx 0.06$).
The effect of the wall perturbation on mixing is even apparent to the eye:
in figure~\ref{fig:mixingresults2}(b), at each of the humps on the outer wall, a shear region could be observed which does not exist in the case of the regular Taylor-Couette geometry.

\subsection{Finite size particles}
In this subsection we study the full cilia-channel-particle problem and compare the results with passive tracers, in an effort to showcase the effects of the particle size in such problems.

\begin{figure}[!h]
        \centerline{\includegraphics[width=.8\linewidth]{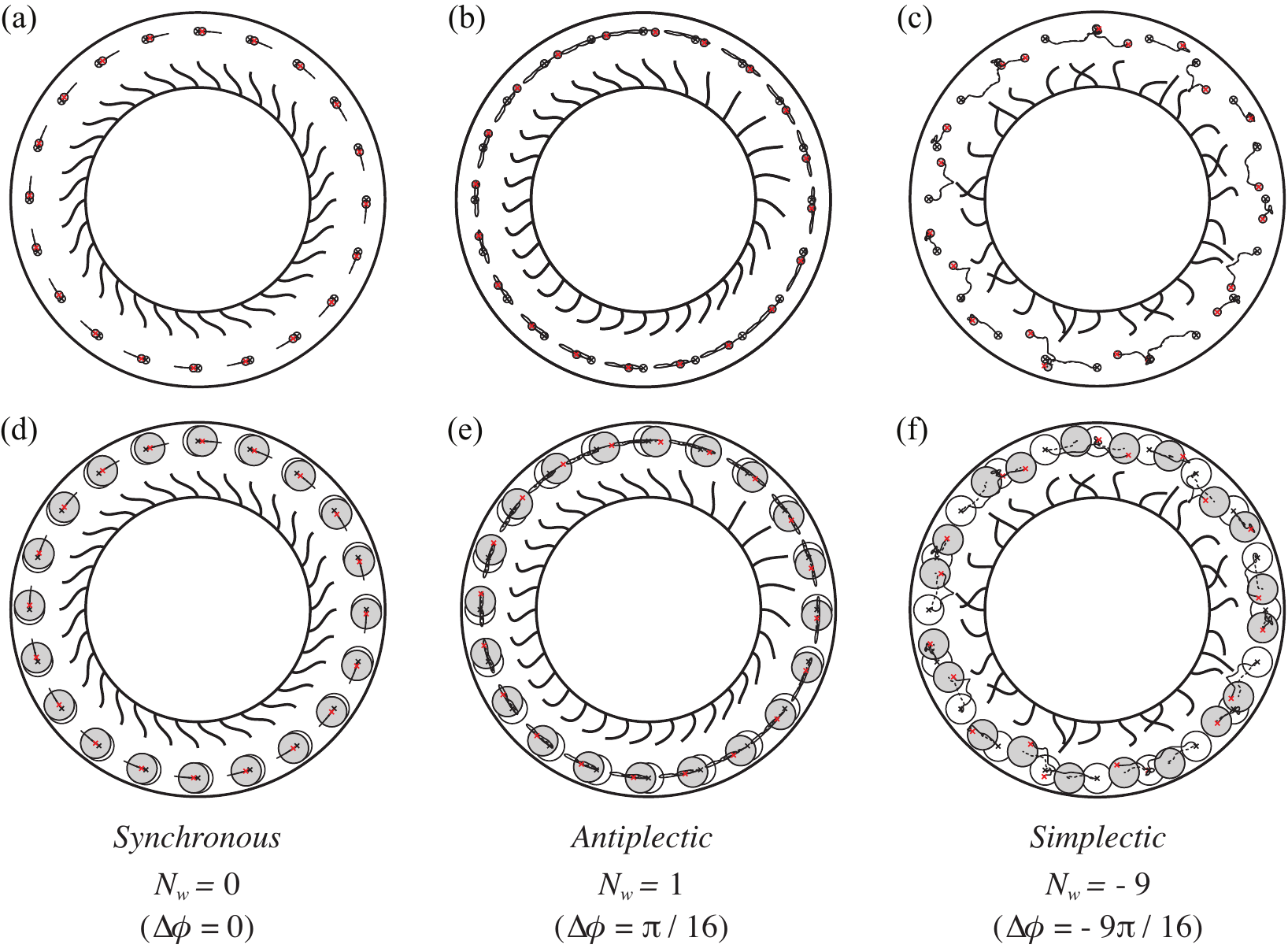}}
	\caption[]{Tracers' trajectories compared to rigid particles' trajectories after one beating cycle.
	Tracers' initial and ending positions are shown in black and red crosses respectively; particles' initial and ending positions are shown in open and solid circles. The trajectories of the tracers and the particle centers are shown in \textcolor{black}{ solid and dashed lines} respectively.
	 (a)-(c): particle radius is $r_p=0.1$; (d)-(f): particle radius is $r_p=0.4$. Left to right: $N_w=0,1,-9$.
	 } 
	\label{fig:cilia-particle-transport}
\end{figure}

We start by uniformly seeding 20 circular particles of radius $r_p$ inside the channel and trace their centroids within one ciliary beating cycle.
With small particle size, as shown in the top row of figure~\ref{fig:cilia-particle-transport}, the differences between the passive tracers and the finite size particles are hardly visible, as expected. 
With large particle size, however, the difference becomes much more evident. 
Specifically, in the first two phase differences ($\Delta\phi = 0, \pi/16$), the motions of the large particles are close to those of the passive tracers, albeit having noticeable shorter distance traveled (\textcolor{black}{dashed} curves have shorter lengths compare to \textcolor{black}{solid} curves).
In the case of large phase difference ($\Delta\phi = -9\pi/16$), the difference between the trajectories of the large particles and the tracers are even more evident.

\begin{figure}[!h]
        \centerline{\includegraphics[width=.9\linewidth]{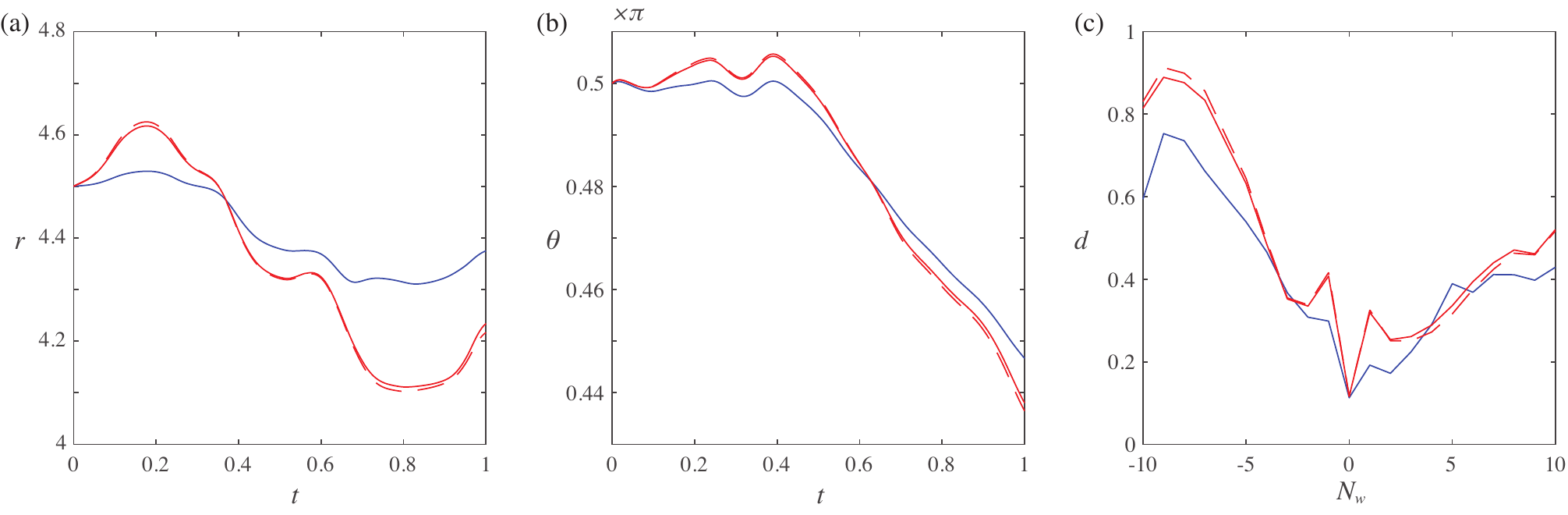}}
	\caption[]{Particle displacement over one beating cycle.
	(a-b) The radial ($r$) and the angular ($\theta$) positions of the particle (tracer) during one beating cycle with $N_w=-9$ ($\Delta\phi = -9\pi/16$).
	(c) The total displacement of the particle (tracer) after one beating cycle as a function of $N_w$. 
	Large ($r_p=0.4$) and small ($r_p = 0.1$) particle results are shown in blue and red lines respectively; tracer result is shown in red dash lines. 
	 } 
	\label{fig:cilia-particle-transport2}
\end{figure}

The radial and azimuthal positions of the tracer and particle initially centered at $(0,4.5)$ are shown in figure~\ref{fig:cilia-particle-transport2}~(a)\&(b).
It is clear that while the differences between the tracer trajectories and small particle trajectories are minimal, large particles deviate from the tracer trajectory since the beginning.
Particularly, the movements of the tracers and the small particles consist of significant deviations in the radial direction, the large particle experiences limited radial deviation throughout the cycle and move in the azimuthal direction only.
The net displacement of the particle over one beating cycle is shown in figure~\ref{fig:cilia-particle-transport2}(c) as a function of $N_w$. 
The net displacement of the large particle is almost always smaller than those of the tracers and the small particles except for two special cases where $N_w=-3$ and $5$. 
Specifically, the net displacement of the particle decreases as much as $40\%$ in the case where $N_w=1$.

\begin{figure}[!h]
        \centerline{\includegraphics[width=.9\linewidth]{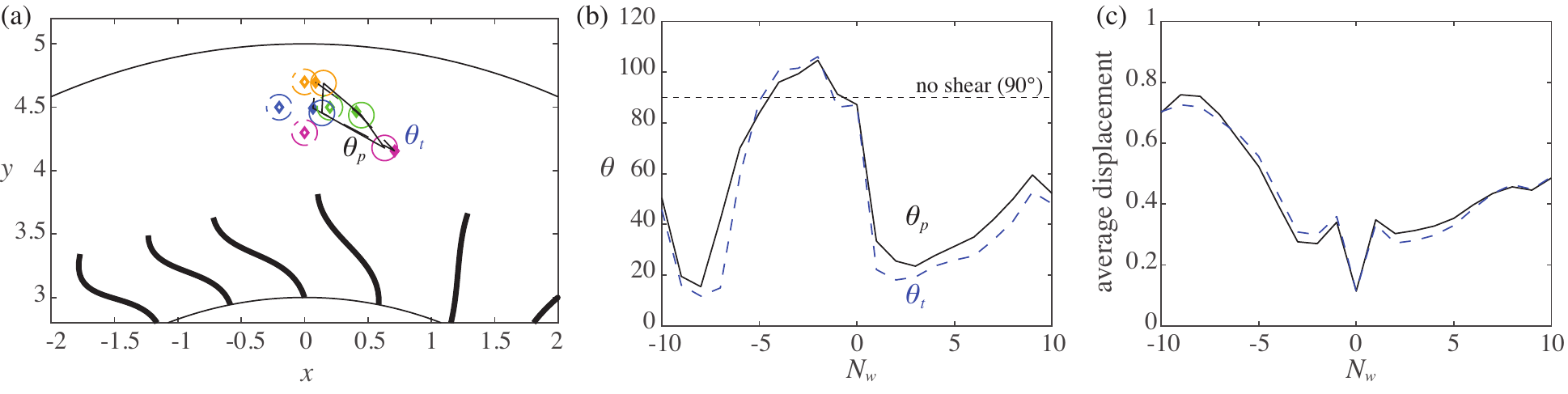}}
	\caption[]{Shear deformation for tracers and rigid particles. Four tracers/particles initially seeded as a square lattice translated by cilia driven flow. (a) A \textcolor{black}{zoomed-in view of the Taylor-Couette device} with $N_w=2$  ($\Delta\phi=2\pi/16$). Initial positions are shown in open diamonds/dashed circles, final positions are shown in closed diamonds/circles. $\theta_t$ and $\theta_p$ denotes the bottom angle of the deformed lattice after one beating cycle for tracers and particles respectively. (b) $\theta_t$ (dashed line) and $\theta_p$ (solid line) as functions of $N_w$. (c) Displacements of the tracers and particles over one cycles, averaged for all tracers (dashed line) and particles (solid line).
	} 
	
	\label{fig:shearresults}
\end{figure}

We conclude this section by studying the effects of the particle sizes in a shear flow. 
In particular, we seed a cluster of 4 particles of radius 0.1 in a square lattice fashion \textcolor{black}{inside the channel. The zoomed-in view is} shown in figure~\ref{fig:shearresults}(a). 
We track the motion of the 4 particles over one beating cycle and measure the angle formed by the bottom 3 particles $\theta_p$ as a metric of shear deformation.
Passive tracers with the same initial positions are also simulated, with the bottom angle denoted by $\theta_t$.
$\theta_p$ and $\theta_t$ are shown in figure~\ref{fig:shearresults}(b) as functions of $N_w$. 
In general, the two angles follow the same trend as we sweep through $N_w$. 
A noticeable fact is that $\theta_p$ is almost always closer to $90^\circ$ compare to $\theta_t$, meaning that the finite-size of the particle is resisting shear deformation when they are in close proximity.
Lastly, we track the displacement of each particle over one cycle and average across all particles as a measure of net transport. The average displacement as a function of $N_w$ is shown in figure~\ref{fig:shearresults}(c). 
Interestingly, having a finite-size does not always result in a smaller or a larger displacement. 
In fact, for most cases where $N_w>0$ or $N_w\le-7$, the cluster of finite-size particles moves farther than the passive tracers; whereas the cluster of passive tracers move farther than particles when $-6\le N_w<0$. 
The differences between the displacements of the particles and the tracers reach 10\% in most cases.

\section{Conclusions and future work}
\label{sc:conclusions}
We presented a hybrid numerical method for simulating cilia-driven particulate flows in complex domains. It features a well-conditioned BIE scheme for handling the moving rigid particles and stationary walls and the method of regularized Stokeslets for handling the cilia. We emphasize that, due to the linearity of Stokes flow, each of the computational modules can be replaced with alternative formulations (e.g., a slender-body theory for handling cilia) or software implementations.    

We applied this method to showcase several examples with varying degree of complexity. In particular, we systematically studied the mixing of fluid tracers inside a ciliary Taylor-Couette device. 
The mixing results are qualitatively different from earlier results obtained using ideal geometries.
Specifically, we demonstrated a case where the mixing region completely dominated the transport region which hasn't been shown before. We also showed that a slight perturbation in the geometry could lead to a global change in the mixing performance: the mixing number increased by about a factor of 2 with the perturbation of the geometry (from $e^{-3.4}\approx 0.03$ to $e^{-2.8}\approx 0.06$). We believe the strong influence of the geometry on the mixing performance is a clear indication that efficient numerical methods that can resolve complex geometry flows are essential for bringing critical insights into ciliary flows observed in natural and engineering applications.
Furthermore, we studied the transport of finite-size particles inside the confining geometry. In particular, we showed that small particles in general behave consistently with fluid tracers (this is a limiting case where the particle size is zero), while large particles impede ciliary-generated movements,  as can be expected. Additionally, we demonstrated that finite-size particles in close proximity resist shear deformation.


We are currently working on extending our work on several fronts. First, we will extend our method to two-way coupled systems, specifically to deformable particles interacting with elastic cilia, capitalizing on prior works such as \cite{Guo2018} and \cite{Veerapaneni2009, rahimian2010dynamic}. \textcolor{black}{Conceptually, our computational scheme can be extended in a straightforward manner to accomodate other two-way coupled models such as \cite{DeCanio2017, Bayly2016, chakrabarti2019spontaneous}. Specifically, in most of the two-way coupled models, the force density along the ciliary centerline, $\boldsymbol{f}$, is a function of configuration or time. Consequently, one can treat $\boldsymbol{f}$ as known and move the related terms  in \eqref{eq:totalvel_mat} to the right-hand-side and solve for $\boldsymbol{\mu}$ only. The ciliary dynamics could then be computed via the no-slip boundary condition~\eqref{eq:bccilia}. We plan to explore these in the near future.} Second, we will consider shape optimization problems, such as optimizing the confining geometry for a given ciliary function (e.g., fluid transport, mixing, etc.) using ideas proposed in~\cite{Bonnet2019}. Extension of this to work to three-dimensional problems is another natural direction.

\begin{acknowledgments}
We thank Shuyang Wang for help with graphics in Figure 1, which uses her REU project work on plotting the streamlines using the line integral convolution technique. We acknowledge support from NSF under grants DMS-1719834 and DMS-1454010.  The work of SV was also supported by the Flatiron Institute, a division of the Simons Foundation. This research was supported in part through computational resources and services provided by the Advanced Research Computing Center and the Mcubed program at the University of Michigan, Ann Arbor.
\end{acknowledgments}
\appendix

\section{Regularized forces} \label{sc:regularized}
The following is the formula for regularized Stokeslet:
\begin{equation}\label{eq:regularizedstokes}
\widetilde{G}_{ij}(\boldsymbol{x},\boldsymbol{y})=\frac{1}{4\pi}\left(\delta_{ij}\left( \log\frac{1}{r_{\epsilon}+\epsilon} -\frac{\epsilon\left(r_{\epsilon}+2\epsilon\right)}{r_{\epsilon}\left(r_{\epsilon}+\epsilon\right)}\right) + (x_i-y_{i})(x_j-y_{j})\dfrac{r_{\epsilon}+2\epsilon}{r_{\epsilon}(r_{\epsilon}+\epsilon)^2}\right),
\end{equation}
where $r_\epsilon=\sqrt{|\boldsymbol{x}-\boldsymbol{y}|^2+\epsilon^2}$, $\epsilon$ is the regularization parameter, and $\delta$ is kronecker delta. Its associated regularized pressure kernel is
\begin{equation}\label{eq:regularizedstokesp}
\begin{split}
\widetilde{P}_{j}(\boldsymbol{x},\boldsymbol{y})=&\frac{\left(x_j-y_j\right)}{2\pi } \frac{r_{\epsilon}^2+\epsilon^2+\epsilon r_{\epsilon}}{ r_{\epsilon}^3\left(r_{\epsilon}+\epsilon\right)}.
\end{split}
\end{equation}

To the best of our knowledge, the traction kernel $\widetilde{T}_{ijk}(\boldsymbol{x},\boldsymbol{y})$ associated with the regularized force has not been given explicitly for the 2D regularized force used in~\cite{Cortez2001}. 
After some lengthy but straightforward derivation following $\widetilde{T}_{ijk}=-\delta_{ij}\widetilde{P}_{k}+\left(\widetilde{G}_{ik,j}+\widetilde{G}_{jk,i}\right)$, we obtain the formula
\begin{equation}
\begin{split}
\widetilde{T}_{ijk}(\boldsymbol{x},\boldsymbol{y}) =& -\frac{(x_i-y_i) (x_j-y_j) (x_k-y_k)}{\pi} 
\frac{r_\epsilon^2+3\epsilon r_\epsilon+\epsilon^2}{r_\epsilon^3 (r_\epsilon+\epsilon)^3} \\
&-  [\delta_{ij}(x_k-y_k)+\delta_{ik}(x_j-y_j)+\delta_{kj}(x_i-y_i)]
\frac{\epsilon^2(2r_\epsilon+ \epsilon)}{2\pi (r_\epsilon+\epsilon)^2 r_\epsilon^3}.
\end{split}
\label{eq:regularized_traction}\end{equation}
It is easy to see that $\widetilde{T}_{ijk}$ converges to the singular traction kernel $T_{ijk}$ in the limit $\epsilon\rightarrow 0$; the correction term induced by the regularization appears at $\mathcal{O}({\epsilon}^2)$ and higher orders of $\epsilon$.

\section{Numerical Validation} \label{sc:validation}
\begin{figure}[!h]
        \centerline{\includegraphics[width=.9\linewidth]{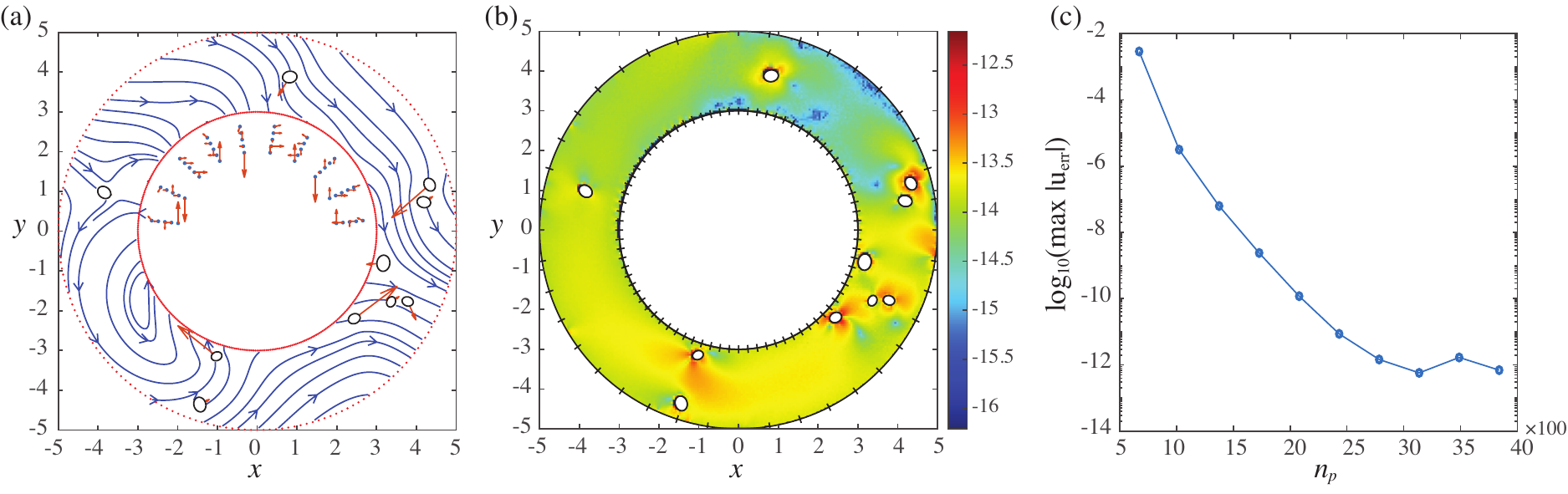}}
	\caption[]{Spatial validation. (a) Flow field generated by 60 stokeslets (red arrows) shown as streamlines. (b) The absolute error between the exact solution and the numerical solution with a total of about 4000 Gaussian quadrature points, color-code represents $\log_{10}(|\boldsymbol{u}_{exa}-\boldsymbol{u}_{num}|)$. (c) The $l_\infty$-norm of the flow field shown as a function of the number of quadrature points.
	} 
	\label{fig:spatialvalidation}
\end{figure}

To validate our boundary integral method, we construct a boundary value problem and test the algorithm against the exact solution. 
Specifically, we place 50 stokeslets with random strengths inside the inner channel boundary and one stokeslet with random strength at the center of each of the 10 particles as shown in figure~\ref{fig:spatialvalidation}(a). 
The flow field $\boldsymbol{u}_{exa}(\Omega)$ created by these stokeslets can be found by evaluating directly using the free-space Green's function.
To obtain the numerical solution, we set the rigid body velocity vector $\boldsymbol{U}^*$  to be zero and treat the flow field on the channel walls and the particle surfaces, given by $\boldsymbol{u}_{exa}({\partial\Omega})$, as the boundary conditions on $\partial\Omega$ where $\partial\Omega\equiv \Gamma\cup\gamma$. 
Symbolically, one can think of $\boldsymbol{u}_{exa}$ as $\boldsymbol{u}^c$ and substitute it into \eqref{eq:fullsys} to solve for the corresponding density function $\boldsymbol\mu$. The numerical solution $\boldsymbol{u}_{num}(\Omega)=\boldsymbol{u}^\Gamma+\boldsymbol{u}^\gamma$ is then found by substituting $\boldsymbol\mu$ into \eqref{eq:boundarydisturb} and \eqref{eq:particldisturb}.

The logarithm of absolute error between $\boldsymbol{u}_{exa}$ and $\boldsymbol{u}_{num}$ is shown in figure~\ref{fig:spatialvalidation}(b) with about 4000 Gauss-Legendre quadrature points on $\Gamma$ and $\gamma$ in total. 
It is noticeable that the algorithm has at least a 14-digit accuracy for most of the locations, and 12-digit accuracy is achieved even close to the particles.
The $l_\infty$-norm of the error as a function of number of quadrature points is shown in figure~\ref{fig:spatialvalidation}(c).

Next, we place $N=32$ cilia with phase difference $\Delta\phi =  2\pi/N = \pi/16$ as in~\eqref{eq:ciliapos} and use a standard Runge-Kutta 4th order (RK4) scheme to march forward in time. Due to the lack of an exact solution in this case, we test the self convergence rate with respect to $\Delta t$. 
We monitor the motion of a rigid particle of radius $r_p=0.4$ initially centered at $(0,4.5)$ for a full cycle $t\in[0,1]$ and for $\Delta t = \{0.04, 0.02, 0.01\}$. 
The particle is discretized using $128$ quadrature points.
At the final time $T=1$, we measure the following quantities in Table~\ref{tab:tmpvalidation}:
\begin{equation}\label{eq:temporalvalidation}
\begin{split}
\mathcal{E}_x(T,\Delta t) &= -\log_2  |x_c^{\Delta t}(T) - x_c^{\Delta t/2}(T)|\\
\mathcal{E}_y(T,\Delta t) &= -\log_2  |y_c^{\Delta t}(T) - y_c^{\Delta t/2}(T)|\\
\mathcal{E}_\Theta(T,\Delta t) &= -\log_2  |\Theta^{\Delta t}(T) - \Theta^{\Delta t/2}(T)|\\
\end{split}
\end{equation}
where $\Theta(t)=\int_0^t \omega \mathrm{d}t$ is the orientation of the particle.
The convergence rate for a passive tracer is also reported in Table~\ref{tab:tmpvalidation}, with $\mathcal{E}_x$ and $\mathcal{E}_y$ only.
\begin{table}
\begin{center}
\begin{tabular}{cccc}
\hline
Particle   & $\Delta t = 0.04$ & $\Delta t = 0.02$& $\Delta t = 0.01$ \\
\hline
$\mathcal{E}_x(T,\Delta t)$ & 13.9948 & 17.0434 & 21.6549 \\
$\mathcal{E}_y(T,\Delta t)$ & 14.6581 & 17.9964 & 22.6611 \\
$\mathcal{E}_\Theta(T,\Delta t)$ & 13.8062 & 17.6920 & 21.6093\\
\hline
\end{tabular}
\hspace{.1in}
\begin{tabular}{cccc}
\hline
Tracer   & $\Delta t = 0.04$ & $\Delta t = 0.02$& $\Delta t = 0.01$ \\
\hline
$\mathcal{E}_x(T,\Delta t)$ & 12.4986 & 16.5527 & 22.1399\\
$\mathcal{E}_y(T,\Delta t)$ & 12.0234 & 16.2576 & 21.3125 \\
 & & & \\
\hline
\end{tabular}
\caption{(Left) Results on the performance of RK4 method applied to evolving the cilia inside a Taylor-Couette device. (Right) Error terms for the particle center at final time $T=1$.}\label{tab:tmpvalidation}
\end{center}
\end{table}


\bibliography{references}

\end{document}